\documentclass[a4paper]{jpconf}%
\usepackage{graphicx}
\usepackage{amsmath}
\usepackage{amsfonts}
\usepackage{amssymb}
\usepackage{graphicx}%
\providecommand{\U}[1]{\protect\rule{.1in}{.1in}}
\begin{document}

\title{The Lee model: a tool to study decays}
\author{Francesco Giacosa$^{\text{(a,b)}}$}

\address{$^{\text{(a)}}$Institute of Physics, Jan Kochanowski University, ul.
Uniwersytecka 7, 25-406 Kielce, Poland}

\address{$^{\text{(b)}}$Institute for Theoretical Physics, Goethe University,
Max-von-Laue-Str.\ 1, 60438 Frankfurt am Main, Germany}

\begin{abstract}
We describe -in a didactical and detailed way- the so-called Lee model (which
shares similarities with the Jaynes-Cummings and Friedrichs models) as a tool
to study unstable quantum states/particles. This Lee model is based on Quantum
Mechanics (QM) but possesses some of the features of Quantum Field Theory
(QFT). The decay process can be studied in great detail and typical\ QFT
quantities such as propagator, one-loop resummation, and Feynman rules can be
introduced. Deviations from the exponential decay law as well as the
Quantum\ Zeno effects can be studied within this framework. The survival
probability amplitude as a Fourier transform of the energy distribution, the
normalization of the latter, and the Breit-Wigner limit can be obtained in a
rigorous mathematical approach.

\end{abstract}

\section{Introduction}

The study of unstable quantum states is a central problem of Quantum Mechanics
(QM) and Quantum Field Theory (QFT) \cite{ghirardi}.

When considering elementary unstable particles or composite unstable hadrons,
QFT represents indeed the correct framework to treat decays. Yet, QFT is
notoriously difficult, and the typical textbook treatment deals with
scattering processes in which `in' and `out' states are regarded as stable
particles (the interaction is eventually switched off in the far past and
future) \cite{peskin} . The discussion of decays (typically the derivation of
the decay width formula) is then performed by generalizing/modifying the
scattering expressions with \textit{ad hoc} arguments. In the textbook of Ref.
\cite{peskin} it is stated that care is needed for the study of decay, since
clearly an unstable state cannot be prepared in the far past.

In this work, we present a useful model -called the Lee model- to treat
unstable states in general and decays in particular. This model, originally
introduced in Ref. \cite{lee}, is rooted in\ QM but -since it contains an
infinite (and countless) number of states- it mimics many features of QFT
\cite{duecan}. The Lee model is exactly solvable, thus offers a very useful
framework to test QFT ideas (it was indeed developed to investigate some
properties of renormalization). Within the Lee model, the time evolution of an
unstable state can be evaluated from an initial time ($t=0$) to any subsequent
time. Quite remarkably, similar models were developed in the literature, such
as the Jaynes-Cummings model used in\ Quantum Optics \cite{jc,jc2,scully} or
the so-called Friedrichs model in mathematical physics
\cite{friedrichs,gadella,horwitz}.

Both in\ QM and in QFT, there is a famous formula that describes the survival
probability amplitude of an unstable state $\left\vert S\right\rangle $
\cite{ghirardi,zenoqft}:
\begin{equation}
a_{S}(t)=\int_{-\infty}^{+\infty}d_{S}(E)e^{-iEt}dE\text{ }.\label{at}%
\end{equation}
Then, the survival probability, i.e. the probability that the state did not
decay yet at the time $t,$ reads $p_{S}(t)=\left\vert a_{S}(t)\right\vert
^{2}.$ Note, when Eq. (\ref{at}) applies, one can show that $a_{S}%
(t\rightarrow\infty)=0,$ i.e.,Poincar\'{e} time is infinite .

The quantity $d_{S}(E)$ is the energy probability distribution for the
unstable state: $d_{S}(E)dE$ is the probability that the energy of the
unstable state $S$ is contained in the interval ($E$,$E+dE$)$.$ A general -and
intuitive- property is its normalization:
\begin{equation}
\int_{-\infty}^{+\infty}d_{S}(E)dE=1.\label{norm}%
\end{equation}
This is quite obvious from a physical point of view, since $p_{S}(0)=1$ (by
construction). Yet its mathematical verification is not trivial.

Two basic and general properties of the function $d_{S}(E)$ have important
consequences: (i) the existence of a low-energy threshold, $d_{S}(E)=0$ for
$E<m_{th}$, implies that for large times $p_{S}(t)$ is a power function
$t^{-\alpha}$ \cite{ghirardi,khalfin,urbanolate}; (ii) the finiteness of the
mean energy, $\left\langle E\right\rangle =\int_{-\infty}^{+\infty}%
Ed_{S}(E)dE<\infty$, implies that $p_{S}^{\prime}(0)=0$ and $p_{S}(t)$ is flat
for short times \cite{ghirardi,duecan,mercouris}. (If $\left\langle
E^{2}\right\rangle $ is also finite, then $p_{S}(t)\simeq1-t^{2}/\tau_{Z}%
^{2},$ as often presented in many papers). Deviations from the exponential law
has been observed at short times by studying the tunneling of sodium atoms in
an optical potential \cite{raizen1}. The experimental verification of
deviations from the exponential law at long times has been obtained by
investigating the fluorescence decays of organic molecules \cite{rothe}.

In this work we intend to prove -in a didactical way and by showing all
intermediate steps- both Eqs. (\ref{at}) and (\ref{norm}) by using the Lee
model mentioned above. Indeed, to show these equations in a pure QFT treatment
is very hard, since one has to go beyond the typical framework used to do QFT
calculations. In this sense, the Lee model offers a clear way to see how QFT
works for decays.

Conversely, the exponential decay law $p_{S}(t)=e^{-\Gamma_{\text{BW}}t}$,
although never exact, is a very good approximation in many physical cases. The
corresponding Breit-Wigner (BW) \cite{ww} energy distribution is:
\begin{equation}
d_{S}^{\text{BW}}(E)=\frac{\Gamma}{2\pi}\left[  (E-M_{\text{BW}})^{2}%
+\Gamma_{\text{BW}}^{2}/4\right]  ^{-1}.\label{bw}%
\end{equation}
Indeed, for many unstable states $d_{S}(E)\simeq d_{S}^{\text{BW}}(E)$ (close
to $M_{\text{BW}}$). Yet, $d_{S}^{\text{BW}}(E)$ is unphysical since it does
not fulfill the two conditions (i) and (ii) mentioned above. By solving the
Lee model we can see how the BW function emerges and how to calculate the
Breit-Wigner mass $M_{\text{BW}}$ and decay width $\Gamma_{\text{BW}}.$

In the end, we also recall that a non-exponential behavior at short times
implies the existence of the so-called Quantum Zeno Effect (QZE) and the
Inverse Zeno Effect (IZE). In the former, the slowing down of the decay due to
frequent measurements takes plac , while in the latter an acceleration of the
decay rate is realized
\cite{dega,misra,facchiprl,shimizu,kk1,kk2,fptopicalreview,fp,gpn}. The
experimental results of Ref. \cite{raizen2}, based on the same setup of Ref.
\cite{raizen1}, could verify also the QZE and IZE effects by adding
intermediate measurements.

The manuscript is organized according to the following scheme:

Section 2: we discuss a simplified version of the problem. We start from the
two-body mixing problem and generalize it obtaining some general formulas in
an heuristic framework.

Section 3: we present the Lee model.\ As a first step the discrete version is
discussed, then the continuous (and correct) Lee Hamiltonian as a limiting
case of the discrete Lee Hamiltonian is presented. Some subtle mathematical
aspects are dealt with care.

Section 4: this section contains the most important results of this work; we
show the validity of Eqs. (\ref{at}) and (\ref{norm}). We shall also present
the Ferynman rules for the Lee model in analogy with QFT. As a last step, we
discuss the Breit-Wigner limit of Eq. (\ref{bw}).

Section 5: we discuss some recent works which made use of the Lee model in
different research topics and summarize their findings.

Section 6: conclusions and outlooks are outlined.

\section{Heuristic presentation of the problem}

In the \textquotedblleft baby version\textquotedblright\ of the problem, let
us first consider two quantum states: a quantum state $\left\vert
S\right\rangle $ (corresponding to the unstable state that we aim to study)
and a quantum state $\left\vert K\right\rangle $ subject to the Hamiltonian
\begin{equation}
H=H_{0}+H_{1}\text{,}\label{hfirst}%
\end{equation}
where $H_{0}$ describes the free (non-interacting) part,
\begin{equation}
H_{0}=M_{S}\left\vert S\right\rangle \left\langle S\right\vert +M_{K}%
\left\vert K\right\rangle \left\langle K\right\vert \text{ }%
\end{equation}
with energies (or masses) $M_{S}$ and $M_{K},$ and $H_{1}$ describes the
`interaction'%
\begin{equation}
H_{1}=g\left(  \left\vert S\right\rangle \left\langle K\right\vert +\left\vert
K\right\rangle \left\langle S\right\vert \right)  \text{ .}%
\end{equation}
In this simple case, the interaction term amounts to a mixing of the two
states, whose strength is controlled by the coupling constant $g.$ In matrix form:%

\begin{equation}
H=\left(
\begin{array}
[c]{cc}%
\left\vert S\right\rangle  & \left\vert K\right\rangle
\end{array}
\right)  \left(
\begin{array}
[c]{cc}%
M_{S} & g\\
g & M_{K}%
\end{array}
\right)  \left(
\begin{array}
[c]{c}%
\left\langle S\right\vert \\
\left\langle K\right\vert
\end{array}
\right)
\end{equation}
where the matrix
\begin{equation}
\Omega=\left(
\begin{array}
[c]{cc}%
M_{S} & g\\
g & M_{K}%
\end{array}
\right)
\end{equation}
has been introduced. The diagonalization of the system is straightforward:
\begin{equation}
H=E_{1}\left\vert E_{1}\right\rangle \left\langle E_{1}\right\vert
+E_{2}\left\vert E_{2}\right\rangle \left\langle E_{2}\right\vert \text{
}\label{hdiag}%
\end{equation}
with
\begin{equation}
\left(
\begin{array}
[c]{c}%
\left\vert E_{1}\right\rangle \\
\left\vert E_{2}\right\rangle
\end{array}
\right)  =\left(
\begin{array}
[c]{cc}%
\cos\theta & \sin\theta\\
-\sin\theta & \cos\theta
\end{array}
\right)  \left(
\begin{array}
[c]{c}%
\left\vert S\right\rangle \\
\left\vert K\right\rangle
\end{array}
\right)
\end{equation}
or
\begin{equation}
\left(
\begin{array}
[c]{c}%
\left\vert S\right\rangle \\
\left\vert K\right\rangle
\end{array}
\right)  =\left(
\begin{array}
[c]{cc}%
\cos\theta & -\sin\theta\\
\sin\theta & \cos\theta
\end{array}
\right)  \left(
\begin{array}
[c]{c}%
\left\vert E_{1}\right\rangle \\
\left\vert E_{2}\right\rangle
\end{array}
\right) \label{o2}%
\end{equation}
Plugging Eq. (\ref{o2}) into Eq. (\ref{hfirst}) and requiring that Eq.
(\ref{hdiag}) emerges leads to the mixing angle as function of $g$ and
masses:
\begin{equation}
\tan2\theta=\frac{-2g}{M_{K}-M_{S}}\text{ .}%
\end{equation}
The energies $E_{1}$ and $E_{2}$ are the eigenvalues of the matrix $\Omega$:
\begin{align}
E_{1} &  =M_{S}\cos^{2}\theta+M_{K}\sin^{2}\theta+g\sin2\theta=\frac{1}%
{2}\left[  M_{0}+M_{K}-\sqrt{\left(  M_{K}-M_{0}\right)  ^{2}+g^{2}}\right]
\text{ ,}\\
E_{2} &  =M_{S}\sin^{2}\theta+M_{K}\cos^{2}\theta-g\sin2\theta=\frac{1}%
{2}\left[  M_{0}+M_{K}+\sqrt{\left(  M_{K}-M_{0}\right)  ^{2}+g^{2}}\right]
\text{ .}%
\end{align}
The state $\left\vert S\right\rangle $ can be written as the superposition of
the energy eigenstates:%
\begin{equation}
\left\vert S\right\rangle =\alpha_{1}\left\vert E_{1}\right\rangle +\alpha
_{2}\left\vert E_{2}\right\rangle
\end{equation}
with $\alpha_{1}=\cos\theta$ and $\alpha_{2}=-\sin\theta.$

Ergo, the \textquotedblleft survival probability amplitude\textquotedblright%
\ for the state $\left\vert S\right\rangle $ is:%
\begin{equation}
a_{S}(t)=\left\langle S\right\vert U(t)\left\vert S\right\rangle =\left\vert
\alpha_{1}\right\vert ^{2}e^{-iE_{1}t}+\left\vert \alpha_{2}\right\vert
^{2}e^{-iE_{2}t}%
\end{equation}
where
\begin{equation}
U(t)=e^{-iHt}%
\end{equation}
is the time-evolution operator. Hence the \textquotedblleft survival
probability\textquotedblright\ is
\begin{equation}
p_{S}(t)=\left\vert a_{S}(t)\right\vert ^{2}.
\end{equation}
It is visible that after the time $T=\frac{2\pi}{E_{2}-E_{1}}$ one has
$p_{S}(T)=1$ (in general $p_{S}(t+T)=p_{S}(t)$). The system is periodic and $T
$ is the Poincar\'{e} recurrence time.

\bigskip

Suppose that now, instead of only two states, we have a $N$-mixing problem,
i.e. we consider the states $\{\left\vert S\right\rangle ,$ $\left\vert
K_{1}\right\rangle ,$ ..., $\left\vert K_{N-1}\right\rangle \}$ with
\begin{equation}
H=M_{S}\left\vert S\right\rangle \left\langle S\right\vert +\sum_{j=1}%
^{N-1}M_{K}\left\vert K_{j}\right\rangle \left\langle K_{j}\right\vert
+g_{j}\left(  \left\vert S\right\rangle \left\langle K_{j}\right\vert
+\left\vert K_{j}\right\rangle \left\langle S\right\vert \right)
\end{equation}
Thus, there are $N-1$ mixing term, each modelled by the own coupling constant
$g_{j}.$

(Note, in principle one could also add the mixing terms proportional to
$\left(  \left\vert K_{i}\right\rangle \left\langle K_{j}\right\vert
+\left\vert K_{j}\right\rangle \left\langle K_{i}\right\vert \right)  ,$ but
this is an unnecessary complication for our purposes.)

By repeating the previous steps one finds the eigenstates $\{\left\vert
E_{1}\right\rangle ,...,\left\vert E_{N}\right\rangle \}$ of the Hamiltonian
$H,$ for which
\begin{equation}
H=\sum_{k=1}^{N}E_{k}\left\vert E_{k}\right\rangle \left\langle E_{k}%
\right\vert \text{ .}%
\end{equation}
Notice that $k=1,...,N$, while in the previous sum $j=1,...,N-1.$

The initial state $\left\vert S\right\rangle $ can be expressed as
\begin{equation}
\left\vert S\right\rangle =\sum_{k=1}^{N}\alpha_{k}\left\vert E_{k}%
\right\rangle
\end{equation}
with the normalization condition
\begin{equation}
\sum_{k=1}^{N}\left\vert \alpha_{k}\right\vert ^{2}=1\text{ .}%
\end{equation}
Then:
\begin{equation}
a_{S}(t)=\left\langle S\right\vert U(t)\left\vert S\right\rangle =\sum
_{k=1}^{N}\left\vert \alpha_{k}\right\vert ^{2}e^{-iE_{k}t}\text{
.}\label{asdisc}%
\end{equation}
Of course, the coefficients $\alpha_{k}$ as well as the energies $E_{k}$ are
function of the parameters of the models: the masses $M_{S}$ and $M_{K,j} $
and the couplings $g_{j}.$ We do not evaluate them here, since this is the
task of the next section. Yet, let us make some simplifying assumptions that
allow to illustrate the problem. We assume that
\begin{equation}
E_{k}=kb\text{ with }b>0\text{ .}%
\end{equation}
The fact that the minimal energy is $b>0$ is an arbitrary choice (one could
anyhow translate the energy to achieve it). In particle physics the minimal
energy corresponds to the sum of the rest masses of the decay process.
Moreover, the maximal energy $E_{\max}$ does not need to be finite. The case
$E_{\max}=\infty$ is indeed pretty common.

Then, in this case, it follows that
\begin{equation}
p_{S}(T=2\pi/b)=1,\text{ }p_{S}(t+T)=p_{S}(t).
\end{equation}
The Poincar\'{e} time $T=2\pi/b$ tends to infinity when $b\rightarrow0.$ Thus,
in this limit, we really have a genuine \textquotedblleft
decay\textquotedblright\ since the original state $\left\vert S\right\rangle $
does not form again at any time $t>0$. The infinite mixing problem implies
that the decay
\begin{equation}
\left\vert S\right\rangle \rightarrow\left\vert K_{j}\right\rangle
\end{equation}
takes place (where of course all $j$ are in principle admitted).

If we decrease $b$ and increase $N$ such that $Nb=E_{\max}$ (by keeping
$E_{\max}$ fixed), the sum of Eq. (\ref{asdisc}) reduces to an integral:%
\begin{align}
a_{S}(t)  &  =\left\langle S\right\vert U(t)\left\vert S\right\rangle
=\sum_{k=1}^{N}\left\vert \alpha_{k}\right\vert ^{2}e^{-iE_{k}t}=\sum
_{k=1}^{N}\frac{\left\vert \alpha_{k}\right\vert ^{2}}{b}be^{-i(kb)t}\\
&  \overset{b\rightarrow0}{=}\int_{0}^{E_{\max}}dmd_{S}(m)e^{-imt}\text{ ,}%
\end{align}
where the continuous variable $m=kb$ has been introduced and the function
$d_{S}(m)$ is given by%
\begin{equation}
d_{S}(m)=\frac{\left\vert \alpha_{k=m/b}\right\vert ^{2}}{b}%
\end{equation}

The normalization condition $\sum_{k=1}^{N}\left\vert \alpha_{k}\right\vert
^{2}$ translates into%
\begin{equation}
\int_{0}^{E_{\max}}dmd_{S}(m)=1.
\end{equation}

Indeed, we have proven in an heuristic way both equations (\ref{at}) and
(\ref{norm}). Even if these arguments do not represent a rigorous proof, they
are intuitive and -as a matter of fact- also correct. Our next task is to
derive them in a formally correct way.

\section{Lee Hamiltonian: definitions and properties}

In this Section we introduce the Lee model. We do it in a two-step process.
First, we consider the discrete case, and then -as a limiting process of the
former the continuous case. The latter represents the final and correct form
of the Lee model.

\subsection{Discrete case}

The basis of the Hilbert space of our problem is assumed to be given by:

\begin{center}
$%
\begin{array}
[c]{c}%
\left\vert S\right\rangle \text{ : the quantum state corresponding to the
unstable state under study,}\\
\left\vert k_{n}\right\rangle \text{ : an infinite set of quantum states
corresponding to decay products.}%
\end{array}
$
\end{center}

The quantum state
\begin{equation}
\left\vert S\right\rangle
\end{equation}
is the state that we aim to investigate. In particular, we will study its time
evolution after its preparation at $t=0.$

The state $\left\vert S\right\rangle $ is not `free' (otherwise the system
would be trivial), but it interacts with other states. To be more precise, it
interacts with an `infinity' of states. We denote these states as
\begin{equation}
\left\vert k_{n}\right\rangle \text{ with }k_{n}=\frac{2n\pi}{L}\text{ and
}n=0,\pm1,\pm2,...\text{ ,}%
\end{equation}
where the quantity $L$ (with the dimension of energy$^{-1}$) can be thought as
the dimension of the linear box in which we place our system. $L$ shall be
regarded as a large number, and indeed in the end the results should not
depend on the box dimension $L$. One already sees that $k_{n}$ can be
interpreted as a `momentum' of the outgoing particles (more details in the
following). Moreover, we consider here only a one-dimensional box: $D=1.$ The
extension to a 3D case is straightforward and -in many physical cases which
embody spherical symmetry- one can reduce a 3D problem into a 1D problem.

Finally, the whole basis of our quantum problem reads:%
\begin{equation}
\text{Basis of the Hilbert space }\mathcal{H}\text{: }\left\{  \left\vert
S\right\rangle ,\left\vert k_{0}\right\rangle ,\left\vert k_{1}\right\rangle
,\left\vert k_{-1}\right\rangle ,...\right\}  \equiv\left\{  \left\vert
S\right\rangle ,\left\vert k_{n}\right\rangle \right\}
\end{equation}
with the usual orthonormal relations:
\begin{equation}
\left\langle S|S\right\rangle =1\text{ , }\left\langle S|k_{n}\right\rangle
=0\text{ },\text{ }\left\langle k_{n}|k_{m}\right\rangle =\delta_{nm}\text{ .}%
\end{equation}
The completeness equation is given by:%
\begin{equation}
\left\vert S\right\rangle \left\langle S\right\vert +\sum_{n}\left\vert
k_{n}\right\rangle \left\langle k_{n}\right\vert =1_{\mathcal{H}}\text{ .}%
\end{equation}

\bigskip

The Hamiltonian of the system consists of two pieces and is constructed in a
similar way as in our `baby' problem of Sec. 2:
\begin{equation}
H=H_{0}+H_{1}\text{ ,}%
\end{equation}
where $H_{0}$ describes the free (non-interacting) part:
\begin{equation}
H_{0}=M_{0}\left\vert S\right\rangle \left\langle S\right\vert +\sum
_{n=0,\pm1,...}\omega(k_{n})\left\vert k_{n}\right\rangle \left\langle
k_{n}\right\vert \text{ ,}%
\end{equation}
and where $H_{1}$ mixes $\left\vert S\right\rangle $ with all $\left\vert
k_{n}\right\rangle $:%
\begin{equation}
H_{1}=\sum_{n=0,\pm1,...}\frac{gf(k_{n})}{\sqrt{L}}\left(  \left\vert
S\right\rangle \left\langle k_{n}\right\vert +\left\vert k_{n}\right\rangle
\left\langle S\right\vert \right)  \text{ .}%
\end{equation}

The following comments are in order:

\begin{itemize}
\item All the coefficients $M_{0},$ $\omega(k_{n}),$ $gf(k_{n})$ are real.

\item The Hamiltonian $H$ is Hermitian.

\item Dimensions: $M_{0}$ and $\omega(k_{n})$ have dimensions [energy], while
$g$ has dimensions [energy$^{+1/2}$]

\item The quantity $M_{0}$ is the bare energy (or mass) of the state
$\left\vert S\right\rangle .$ Note, introduce the `bare mass' $M_{0}$ (instead
$M_{S}$) since a dressing process takes place and the mass of the state $S$ is
in general shifted by quantum fluctuations.

\item The energy $\omega(k_{n})$ is the (bare) energy of the state $\left\vert
k_{n}\right\rangle .$

\item The coupling constant $g$ measures the strength of the interaction; the
(dimensionless) form factor $f(k_{n})$ modulates the interaction. In practice,
each mixing $\left\vert S\right\rangle \longleftrightarrow\left\vert
k_{n}\right\rangle $ has its own coupling constant $gf(k_{n}).$

\item The factor$\sqrt{L}$ is introduced for future convenience: it is
necessary for a smooth continuous limit in which $L\rightarrow\infty.$

\item For simplicity of notations, $\sum_{n=0,\pm1,...}$can be also expressed
as $\sum_{n}$ (or as $\sum_{k}$).
\end{itemize}

\bigskip

It is also important to discuss the physical interpretation of the set-up.
Thinking in terms of unstable particles, the state $\left\vert S\right\rangle
$ represents an unstable particle $S$ in its rest frame. That means, the total
momentum of $\left\vert S\right\rangle $ vanishes. The state $\left\vert
k_{n}\right\rangle $ represents a possible final state of the decay of $S.$ In
the simplest case of a two-body decay, the state $\left\vert k_{n}%
\right\rangle $ represents \textbf{two} particles emitted by $S$ and flying
back-to-back. What we have in mind is a decay of the type%
\begin{equation}
S\rightarrow\varphi_{1}+\varphi_{2}\text{ .}%
\end{equation}

In the case of a spacial one-dimensional decay, $k_{n}$ can be interpreted as
the momentum of the first emitted particle, while $-k_{n}$ is the momentum of
the second emitted particle. Schematically:
\begin{equation}
\left\vert k_{n}\right\rangle \equiv\left\vert \varphi_{1}(k_{n}),\varphi
_{2}(-k_{n})\right\rangle
\end{equation}
In this way, the total three-momentum of $\left\vert k_{n}\right\rangle $ is
still zero, as it must. The 3D extension is straightforward.

As possible and clarifying examples of such a process we mention:

(i) The neutral pion $\pi^{0}$ decays into two photons: $\pi^{0}%
\rightarrow\gamma\gamma.$ Then, $\pi^{0}$ in its rest frame corresponds to
$\left\vert S\right\rangle ,$ while $\gamma\gamma$ corresponds to $\left\vert
k_{n}\right\rangle $ (one photon has momentum $k_{n},$ the other $-k_{n}$).

(ii) An excited atom $A^{\ast}$ decays into the-ground state atom $A$ emitting
a photon $\gamma$: $A^{\ast}\rightarrow A\gamma.$ In this case, $A^{\ast}$ is
the sate $\left\vert S\right\rangle ,$ while $\left\vert k_{n}\right\rangle $
represents the joint system of the ground-state atom $A$ and the photon.

Clearly, a huge numbers of such examples can be presented. In general, it is
not necessary to consider only a two-body decay. It is just simpler doing so
for obvious reasons. Yet, the important point is that there is an infinity of
states of the type $\left\vert k_{n}\right\rangle $, one for each $k_{n}=2\pi
n/L$.

While the Lee Hamiltonian has its own validity even without the present
analogy to particle physics, it should be actually stressed that this is more
then an analogy. One can namely show that a Quantum\ Field Theory (under
certain approximations) reduces to a Lee Hamiltonian \cite{duecan}.

\bigskip

\textbf{Function }$\omega(k_{n})$: the function $\omega(k_{n})$ represents the
energy of the state $\left\vert k_{n}\right\rangle .$ In the case of a
two-body particle decay, its form is given by%
\begin{equation}
\omega(k_{n})=\sqrt{k_{n}^{2}+m_{1}^{2}}+\sqrt{k_{n}^{2}+m_{2}^{2}}\text{
,}\label{omega}%
\end{equation}
where $m_{1}$ is the mass of the first particle $\varphi_{1}$ and $m_{2}$ of
the second particle $\varphi_{2}.$ Clearly:
\begin{equation}
\omega(k_{n})\geq m_{1}+m_{2}\text{ .}%
\end{equation}
In the two-photon decay such as the process (i) described above, one has
$m_{1}=m_{2}=0,$ hence%
\begin{equation}
\omega(k_{n})=2\left\vert k_{n}\right\vert .
\end{equation}
In the atomic decay (ii) described above, one has $M_{0}=M_{A}+\Delta,$
$m_{1}=0,$ and $m_{2}=M_{A},$ hence:%
\begin{equation}
\omega(k_{n})\simeq\left\vert k_{n}\right\vert +M_{A}.
\end{equation}
In this case, one could also subtract a constant term, $H\rightarrow
H-M_{A}1_{\mathcal{H}}.$ Hence,
\begin{equation}
\omega(k_{n})\simeq\left\vert k_{n}\right\vert \label{lin}%
\end{equation}
holds.

\bigskip

\textbf{Linear Lee Model (LLM): }A useful model, that we call `Linear Lee
model' (LLM) is obtained by making the choice
\begin{equation}
\omega(k_{n})=k_{n}\text{ ,}\label{linearleemodel}%
\end{equation}
in which the energy function of the state $\left\vert k_{n}\right\rangle $ has
a simple linear form.

Clearly, the fact that negative values of the energy are admitted makes it
fundamentally different from Eq. (\ref{omega}). Namely, for $\omega
(k_{n})=k_{n}$ there is no minimal energy of the system, which is clearly an
unphysical property.

Even when the masses $m_{1}$ and $m_{2}$ vanishes, one has $\omega
(k_{n})=2\left\vert k_{n}\right\vert ,$ where the modulus is present. Also, we
may consider the limiting case of Eq. (\ref{lin}), $\omega(k_{n}%
)\simeq\left\vert k_{n}\right\vert $ where -again- the modulus is present.
Yet, we might imagine the photon is always emitted to the \textquotedblleft
right\textquotedblright, hence $\omega(k_{n})\simeq k_{n}>0.$ In this respect,
Eq. (\ref{linearleemodel}) would represent a good (sometimes extremely good)
numerically approximation to this problem, but from a fundamental point of
view we still violate basic properties of our original system.

Yet, the LLM has some advantages:

(i) It allows in most cases for simple analytic expressions. The corresponding
properties are similar also in the case of an arbitrary (and more realistic)
function $\omega(k).$

(ii) The exponential limit for the decay of $S$ can be nicely obtained for
$f(k)=1$, see details later.

(iii) The LLM is nevertheless not trivial. For an arbitrary $f(k),$ one has
the all the nontrivial properties that one expects: nonexponential decay law
for short and long times as well as non-trivial scattering of the type
$\left\vert k_{n}\right\rangle \rightarrow$ $\left\vert k_{m}\right\rangle .$
In this respect, the LLM model is not a defined model as long as $f(k)$ is not
determined. Even if extremely simplified, it describes an infinity of possible models.

\subsection{Continuous case}

Let us now turn to the continuous Lee model. To this end, we perform the limit
$L\rightarrow\infty$, implying that the variable $k_{n}$ becomes continuos:
\begin{equation}
k_{n}=\frac{2\pi n}{L}\rightarrow k\in(-\infty,+\infty)\text{ }.
\end{equation}
This limit is however rather subtle and requires some care. When $L$ is sent
to infinity, the sum turns to an integral:%
\begin{equation}
\sum_{n}=\frac{L}{2\pi}\sum_{n}\frac{2\pi}{L}\rightarrow\frac{L}{2\pi}%
\int_{-\infty}^{+\infty}dk=L\int_{-\infty}^{+\infty}\frac{dk}{2\pi}\text{ }%
\end{equation}
where $\delta k=L/2\pi$ has been introduced in order to generate the
differential $dk.$

Yet, the subtle piece is to note that the kets must change. Namely, in terms
of continuous variables we expect that%
\begin{equation}
\left\langle k_{1}|k_{2}\right\rangle =\delta(k_{1}-k_{2})\text{. }%
\end{equation}
To this end, let us write down the following $L$-dependent representation of
the Dirac-delta function:%

\begin{equation}
\delta_{L}(k_{n})=\left\{
\begin{array}
[c]{c}%
0\text{ for }n\neq0\\
\frac{L}{2\pi}\text{for }n=0
\end{array}
\right.  \text{ .}%
\end{equation}
Clearly:
\begin{equation}
\delta_{L}(0)=\frac{L}{2\pi}\text{ .}%
\end{equation}
As a proof that $\delta_{L}(k_{n})$ has the desired properties we evaluate
\begin{equation}
\sum_{n}\delta k\delta_{L}(k_{n})=1\text{ }\forall L\rightarrow\sum_{n}\delta
k\delta_{L}(k_{n})u(k_{n})=u(0)\text{ ,}%
\end{equation}
as it should. Hence, in the limit $L\rightarrow\infty$ one obtains:
\begin{equation}
\sum_{n}\delta k\delta_{L}(k_{n})u(k_{n})\rightarrow\int_{-\infty}^{+\infty
}dk\delta(k)u(k)=u(0)\text{ ,}%
\end{equation}
showing that we have obtained a representation of the Dirac function as:
\begin{equation}
\delta(k)=\lim_{L\rightarrow\infty}\delta_{L}(k_{n})\text{ }.
\end{equation}
We can verify the results also by using the standard integral representation%
\begin{equation}
\delta_{L}(k_{n})=\int_{-L/2}^{L/2}\frac{dx}{2\pi}e^{ik_{n}x}=\left\{
\begin{array}
[c]{c}%
0\text{ for }n\neq0\\
\frac{L}{2\pi}\text{for }n=0
\end{array}
\right.  \text{ .}%
\end{equation}

Finally, the link between $\left\vert k_{n}\right\rangle $ and $\left\vert
k\right\rangle $ is given by:%
\begin{equation}
\left\vert k_{n}\right\rangle =\sqrt{\frac{2\pi}{L}}\left\vert k\right\rangle
\text{ .}%
\end{equation}
In fact, in this way:
\begin{equation}
\left\langle k_{1}|k_{2}\right\rangle =\lim_{L\rightarrow\infty}\frac{L}{2\pi
}\left\langle k_{n_{1}}|k_{n_{2}}\right\rangle =\lim_{L\rightarrow\infty
}\left\{
\begin{array}
[c]{c}%
0\text{ for }n_{1}\neq n_{2}\\
\frac{L}{2\pi}=\delta_{L}(0)\text{ for }n_{1}=n_{2}%
\end{array}
\right.  =\delta(k_{1}-k_{2})
\end{equation}
as expected.

[Note, the extension to $D=3$ is straightforward:
\begin{equation}
\sum_{\mathbf{k}}\rightarrow V\int\frac{d^{3}k}{(2\pi)^{3}}\text{ ,}%
\end{equation}
where
\begin{equation}
V=L^{3}\text{ and }\left\vert \vec{k}_{discrete}\right\rangle \rightarrow
(2\pi)^{3/2}/\sqrt{V}\left\vert \vec{k}_{cont}\right\rangle
\end{equation}
is the link between discrete and continuous kets.]

It is also quite peculiar that the very dimension of the ket has changed in
the passage:%
\begin{equation}%
\begin{array}
[c]{c}%
\dim[\left\vert k_{n}\right\rangle ]=[\text{Energy}^{0}]\text{
(dimensionless)}\\
\dim[\left\vert k\right\rangle ]=[\text{Energy}^{-1/2}]\text{ }%
\end{array}
\text{ .}%
\end{equation}
Then, the continuos Hilbert space is given by:%
\begin{equation}
\mathcal{H}=\left\{  \left\vert S\right\rangle ,\left\vert k\right\rangle
\right\}
\end{equation}
with%
\begin{equation}
\left\langle S|S\right\rangle =1\text{ , }\left\langle S|k\right\rangle
=0\text{ },\text{ }\left\langle k_{1}|k_{2}\right\rangle =\delta(k_{1}%
-k_{2})\text{ .}%
\end{equation}
We also check the completeness relation:
\begin{align}
1_{\mathcal{H}}  &  =\left\vert S\right\rangle \left\langle S\right\vert
+\sum_{n}\left\vert k_{n}\right\rangle \left\langle k_{n}\right\vert
=\left\vert S\right\rangle \left\langle S\right\vert +\sum_{n}\delta k\left(
\sqrt{\frac{L}{2\pi}}\left\vert k_{n}\right\rangle \left\langle k_{n}%
\right\vert \sqrt{\frac{L}{2\pi}}\right) \\
&  \rightarrow\left\vert S\right\rangle \left\langle S\right\vert
+\int_{-\infty}^{+\infty}dk\left\vert k\right\rangle \left\langle k\right\vert
=1_{\mathcal{H}}\text{ .}\label{unity}%
\end{align}

Finally, we are ready to present the Lee Hamiltonian in the continuous limit:%

\begin{equation}
H=H_{0}+H_{1}%
\end{equation}
where%

\begin{align}
H_{0} &  =M\left\vert S\right\rangle \left\langle S\right\vert +\int_{-\infty
}^{+\infty}dk\omega(k)\left\vert k\right\rangle \left\langle k\right\vert
\text{ , }\nonumber\\
H_{1} &  =\int_{-\infty}^{+\infty}dk\frac{gf(k)}{\sqrt{2\pi}}\left(
\left\vert S\right\rangle \left\langle k\right\vert +\left\vert k\right\rangle
\left\langle S\right\vert \right)  \text{ .}\label{contlee}%
\end{align}
One can verify that the dimensions is preserved. For instance:
\begin{align}
\dim\left[  dk\omega(k)\left\vert k\right\rangle \left\langle k\right\vert
\right]   &  =\dim\left[  dk]\dim[\omega(k)]\dim^{2}[\left\vert k\right\rangle
\right] \\
&  =[\text{Energy}][\text{Energy}][\text{Energy}^{-1}]=[\text{Energy}]
\end{align}
[The continuous LLM is obtained upon setting $\omega(k)=k$.]

The framework for the study of the time evolution is ready.

\section{Survival amplitude, propagator, and spectral function}

\subsection{Time-evolution operator}

The Schr\"{o}dinger equation (in natural units)
\begin{equation}
i\frac{\partial\left\vert \psi(t)\right\rangle }{\partial t}=H\left\vert
\psi(t)\right\rangle
\end{equation}
can be univocally solved for a certain given initial state
\begin{equation}
\left\vert \psi(0)\right\rangle =\beta_{S}\left\vert S\right\rangle +\sum
_{n}\beta_{n}\left\vert k_{n}\right\rangle \overset{L\rightarrow\infty}%
{\equiv}\beta_{S}\left\vert S\right\rangle +\int_{-\infty}^{+\infty}%
dk\beta(k)\left\vert k\right\rangle
\end{equation}
with:
\begin{equation}
\beta(k)\overset{L\rightarrow\infty}{\equiv}\sqrt{\frac{L}{2\pi}}\beta_{n}%
\end{equation}
Hence:%
\begin{equation}
1=\left\vert \beta_{S}\right\vert ^{2}+\sum_{n}\left\vert \beta_{n}\right\vert
^{2}\overset{L\rightarrow\infty}{\equiv}\left\vert \beta_{S}\right\vert
^{2}+\int_{-\infty}^{+\infty}dk\left\vert \beta(k)\right\vert ^{2}\text{ }%
\end{equation}
In particular, we shall be interested to the case $\beta_{S}=1$ and $\beta
_{n}\equiv\beta(k)=0$.

A formal solution to the time evolution is obtained by introducing the
time-evolution operator:%

\begin{equation}
U(t)=e^{-iHt}\text{ }%
\end{equation}
out of which%
\begin{equation}
\left\vert \psi(t)\right\rangle =U(t)\left\vert \psi(0)\right\rangle \text{ .
}%
\end{equation}

The time-evolution operator can be expressed in terms of a Fourier transform:%
\begin{equation}
U(t)=\frac{i}{2\pi}\int_{-\infty}^{+\infty}dE\frac{1}{E-H+i\varepsilon
}e^{-iEt}\text{ }=\frac{i}{2\pi}\int_{-\infty}^{+\infty}dEG(E)e^{-iEt}\text{
(for }t>0\text{)}\label{uprop}%
\end{equation}
where $\varepsilon$ is an infinitesimal number and where the `propagator
operator'
\begin{equation}
G(E)=\frac{1}{E-H+i\varepsilon}%
\end{equation}
has been introduced. For $t>0$ one can formally close the integral in the
lower half complex plane. In fact:
\begin{equation}
e^{-iEt}=e^{-it\operatorname{Re}E}e^{t\operatorname{Im}E}%
\end{equation}
means that for $t>0$ one should consider $\operatorname{Im}E<0$ in such a way
that the $e^{-iEt}$ goes to zero when $\left\vert E\right\vert \rightarrow
\infty.$ Then, the residue theorem assures that Eq. (\ref{uprop}) is correct.
This equality holds also at the level of operators since it is valid for any
eigenstate of $H$ (see the next section for an explicit example).

The propagator operator $G(E)$ can be expanded as:%

\begin{align}
G(E)  &  =\frac{1}{E-H+i\varepsilon}=\frac{1}{E-H_{0}-H_{1}+i\varepsilon
}\nonumber\\
&  =\frac{1}{\left(  E-H_{0}+i\varepsilon\right)  \left(  1-\frac{1}%
{E-H_{0}+i\varepsilon}H_{1}\right)  }\nonumber\\
&  =\frac{1}{\left(  1-\frac{1}{E-H_{0}+i\varepsilon}H_{1}\right)  }\frac
{1}{\left(  E-H_{0}+i\varepsilon\right)  }\nonumber\\
&  =\sum_{n=0}^{\infty}\left(  \frac{1}{E-H_{0}+i\varepsilon}H_{1}\right)
^{n}\frac{1}{E-H_{0}+i\varepsilon}\label{propexp}%
\end{align}
where we have used that $(AB)^{-1}=B^{-1}A^{-1}$ ($A,B$ arbitrary operators on
$\mathcal{H}$).

\subsection{Survival probability's amplitude of $S$}

We are interested in the evaluation of the survival probability amplitude%
\begin{equation}
a_{S}(t)=\left\langle S\right\vert U(t)\left\vert S\right\rangle \text{ .}%
\end{equation}
out of which the survival probability of the state $S$ reads:%
\begin{equation}
p_{S}(t)=\left\vert a_{S}(t)\right\vert ^{2}.
\end{equation}

\bigskip

\textit{Trivial limit}: Let us first consider the a trivial example: $H=H_{0}
$ (this corresponds to the limit $g\rightarrow0$, no interaction and no decay).

Way 1:
\begin{align}
a_{S}(t)  &  =\left\langle S\right\vert U(t)\left\vert S\right\rangle
=\left\langle S\right\vert e^{-iH_{0}t}\left\vert S\right\rangle =e^{-iM_{0}%
t}\text{ }\\
&  \rightarrow p_{S}(t)=1\text{ (stable state).}%
\end{align}

Way 2:
\begin{equation}
a_{S}(t)=\left\langle S\right\vert U(t)\left\vert S\right\rangle =\left\langle
S\right\vert \frac{i}{2\pi}\int_{-\infty}^{+\infty}dE\frac{1}{E-H_{0}%
+i\varepsilon}e^{-iEt}\left\vert S\right\rangle =\frac{i}{2\pi}\int_{-\infty
}^{+\infty}dE\frac{1}{E-M_{0}+i\varepsilon}e^{-iEt}\text{ .}%
\end{equation}
The latter integral can be solved by using the Jordan lemma (close down and
pick up the pole for $E=M_{0}-i\varepsilon.$ Note, as discussed above, one is
obliged to close down for $t>0$). Then, by using the residue theorem we
obtain:%
\begin{equation}
a_{S}(t)=\frac{i}{2\pi}(-1)2\pi ie^{-i(M_{0}-\varepsilon)t}\text{ ,}%
\end{equation}
where the extra-factor $(-1)$ comes from the fact that the path is followed
clockwise. Finally, by sending $\varepsilon\rightarrow0$ we get the expected
result:%
\begin{equation}
a_{S}(t)=e^{-iM_{0}t}\text{ }\rightarrow p_{S}(t)=1\text{ .}%
\end{equation}
In passing by, we note that the object
\begin{equation}
G_{S}^{\text{free}}(E)=G_{S}^{(0)}(E)=\left\langle S\right\vert \frac
{1}{E-H_{0}+i\varepsilon}\left\vert S\right\rangle =\frac{1}{E-M_{0}%
+i\varepsilon}%
\end{equation}
is called the free propagator of the state $S.$

\bigskip

\textit{Evaluation of }$a(t)$\textit{\ in the full case. }In the full case one
proceeds as follow. The survival amplitude $a_{S}(t)$ takes the form%
\begin{equation}
a_{S}(t)=\frac{i}{2\pi}\int_{-\infty}^{+\infty}dEG_{S}(E)e^{-iEt}\text{ }%
\end{equation}
where%
\begin{equation}
G_{S}(E)=\left\langle S\right\vert G(E)\left\vert S\right\rangle =\left\langle
S\right\vert \frac{1}{E-H+i\varepsilon}\left\vert S\right\rangle
\end{equation}
is the full propagator of $S.$

It is now necessary to evaluate $G_{S}(E)$ explicitly. As a first step, we use
Eq. (\ref{propexp}) obtaining the expansion:%
\begin{equation}
G_{S}(E)=\left\langle S\right\vert G(E)\left\vert S\right\rangle =\sum
_{n=0}^{\infty}\left\langle S\right\vert \left(  \frac{1}{E-H_{0}%
+i\varepsilon}H_{1}\right)  ^{n}\frac{1}{E-H_{0}+i\varepsilon}\left\vert
S\right\rangle =\sum_{n=0}^{\infty}G_{S}^{(n)}(E)
\end{equation}
with%
\begin{equation}
G_{S}^{(n)}(E)=\left\langle S\right\vert \left(  \frac{1}{E-H_{0}%
+i\varepsilon}H_{1}\right)  ^{n}\left\vert S\right\rangle \frac{1}%
{E-M_{0}+i\varepsilon}\text{ .}%
\end{equation}
Let us evaluate the first three terms:%
\begin{equation}
n=0\rightarrow G_{S}^{(0)}(E)=\left\langle S\right\vert 1\left\vert
S\right\rangle \frac{1}{E-M_{0}+i\varepsilon}=\frac{1}{E-M_{0}+i\varepsilon
}\text{ ,}%
\end{equation}%
\begin{equation}
n=1\rightarrow\left\langle S\right\vert \frac{1}{E-H_{0}+i\varepsilon}%
H_{1}\left\vert S\right\rangle \frac{1}{E-M_{0}+i\varepsilon}=0\text{ ,}%
\end{equation}

\begin{align}
n &  =2\rightarrow G_{S}^{(1)}(E)=\left\langle S\right\vert \left(  \frac
{1}{E-H_{0}+i\varepsilon}H_{1}\right)  ^{2}\left\vert S\right\rangle \frac
{1}{E-M_{0}+i\varepsilon}\\
&  =\frac{1}{E-M_{0}+i\varepsilon}\left\langle S\right\vert H_{1}\frac
{1}{E-H_{0}+i\varepsilon}H_{1}\left\vert S\right\rangle \frac{1}%
{E-M_{0}+i\varepsilon}\\
&  =-\frac{\Pi(E)}{\left(  E-M_{0}+i\varepsilon\right)  ^{2}}\text{ .}%
\end{align}
The recursive quantity is $\Pi(E)$:%
\begin{equation}
\Pi(E)=-\left\langle S\right\vert H_{1}\frac{1}{E-H_{0}+i\varepsilon}%
H_{1}\left\vert S\right\rangle \text{ .}%
\end{equation}
We introduce $1_{\mathcal{H}}=\left\vert S\right\rangle \left\langle
S\right\vert +\int_{-\infty}^{+\infty}dk\left\vert k\right\rangle \left\langle
k\right\vert $ two times, obtaining:
\begin{align}
\Pi(E) &  =-\left\langle S\right\vert H_{1}1_{\mathcal{H}}\frac{1}%
{E-H_{0}+i\varepsilon}1_{\mathcal{H}}H_{1}\left\vert S\right\rangle
=\nonumber\\
&  =-\int_{-\infty}^{+\infty}dk\int_{-\infty}^{+\infty}dq\left\langle
S\right\vert H_{1}\left\vert k\right\rangle \left\langle k\right\vert \frac
{1}{E-H_{0}+i\varepsilon}\left\vert q\right\rangle \left\langle q\right\vert
H_{1}\left\vert S\right\rangle \\
&  =-\int_{-\infty}^{+\infty}dk\int_{-\infty}^{+\infty}dq\frac{gf(k)}%
{\sqrt{2\pi}}\frac{\delta(k-q)}{E-\omega(k)+i\varepsilon}\frac{gf(q)}%
{\sqrt{2\pi}}\nonumber\\
&  =-\int_{-\infty}^{+\infty}\frac{dk}{2\pi}\frac{g^{2}f(k)^{2}}%
{E-\omega(k)+i\varepsilon}=g^{2}\Sigma(E)\text{ .}%
\end{align}
where we have used that%
\begin{equation}
\left\langle S\right\vert H_{1}\left\vert k\right\rangle =\frac{gf(k)}%
{\sqrt{2\pi}}\text{ .}%
\end{equation}
Summarizing:
\begin{equation}
\Pi(E)=g^{2}\Sigma(E)=-\int_{-\infty}^{+\infty}\frac{dk}{2\pi}\frac
{g^{2}f(k)^{2}}{E-\omega(k)+i\varepsilon}%
\end{equation}
Going further, we get:%
\begin{align}
G_{S}^{(3)}(E) &  =\left\langle S\right\vert \left(  \frac{1}{E-H_{0}%
+i\varepsilon}H_{1}\right)  ^{3}\left\vert S\right\rangle \frac{1}%
{E-M_{0}+i\varepsilon}\\
&  =\frac{1}{E-M_{0}+i\varepsilon}\left\langle S\right\vert H_{1}\frac
{1}{E-H_{0}+i\varepsilon}H_{1}\frac{1}{E-H_{0}+i\varepsilon}H_{1}\left\vert
S\right\rangle \frac{1}{E-M_{0}+i\varepsilon}\\
&  =0
\end{align}
Namely, one can again insert $1_{\mathcal{H}}$, but an odd number of $H_{1}$
implies that this amplitude vanishes. In general:
\begin{equation}
G_{S}^{(2n+1)}(E)=0\text{, }n=0,1,2,..
\end{equation}
Next, by properly inserting $1_{\mathcal{H}}$ two times:%
\begin{align}
G_{S}^{(4)}(E) &  =\left\langle S\right\vert \left(  \frac{1}{E-H_{0}%
+i\varepsilon}H_{1}\right)  ^{4}\left\vert S\right\rangle \frac{1}%
{E-M_{0}+i\varepsilon}\nonumber\\
&  =\frac{1}{E-M_{0}+i\varepsilon}\left\langle S\right\vert H_{1}\frac
{1}{E-H_{0}+i\varepsilon}H_{1}\frac{1}{E-H_{0}+i\varepsilon}H_{1}\frac
{1}{E-H_{0}+i\varepsilon}H_{1}\left\vert S\right\rangle \frac{1}%
{E-M_{0}+i\varepsilon}\nonumber\\
&  \frac{1}{E-M_{0}+i\varepsilon}\left\langle S\right\vert H_{1}\frac
{1}{E-H_{0}+i\varepsilon}H_{1}1_{\mathcal{H}}\frac{1}{E-H_{0}+i\varepsilon
}1_{\mathcal{H}}H_{1}\frac{1}{E-H_{0}+i\varepsilon}H_{1}\left\vert
S\right\rangle \frac{1}{E-M_{0}+i\varepsilon}\nonumber\\
&  =\frac{1}{\left(  E-M_{0}+i\varepsilon\right)  ^{2}}\left[  \int_{-\infty
}^{+\infty}dk_{1}\frac{gf(k_{1})}{\sqrt{2\pi}}\frac{1}{E-\omega(k_{1}%
)+i\varepsilon}\frac{gf(k_{1})}{\sqrt{2\pi}}\right]  \frac{1}{E-M_{0}%
+i\varepsilon}\nonumber\\
&  \times\left[  \int_{-\infty}^{+\infty}dk_{2}\frac{gf(k_{2})}{\sqrt{2\pi}%
}\frac{1}{E-\omega(k_{2})+i\varepsilon}\frac{gf(k_{2})}{\sqrt{2\pi}}\right]
\frac{1}{E-M_{0}+i\varepsilon}\nonumber\\
&  =\frac{\Pi(E)^{2}}{\left(  E-M_{0}+i\varepsilon\right)  ^{3}}\text{ .}%
\end{align}
Putting all the pieces together:%
\begin{equation}
G_{S}^{(2n)}(E)=\frac{\left[  -\Pi(E)\right]  ^{n}}{\left(  E-M_{0}%
+i\varepsilon\right)  ^{n+1}}\text{ .}%
\end{equation}
Finally:%
\begin{align}
G_{S}(E) &  =\sum_{n=0}^{\infty}G_{S}^{(n)}(E)=\sum_{n=0}^{\infty}G_{S}%
^{(2n)}(E)=\sum_{n=0}^{\infty}\frac{\left[  -\Pi(E)\right]  ^{n}}{\left(
E-M_{0}+i\varepsilon\right)  ^{2n+1}}\nonumber\\
&  =\frac{1}{\left(  E-M_{0}+i\varepsilon\right)  }\sum_{n=0}^{\infty}%
\frac{\left[  -\Pi(E)\right]  ^{n}}{\left(  E-M_{0}+i\varepsilon\right)  ^{n}%
}=\frac{1}{\left(  E-M_{0}+i\varepsilon\right)  }\frac{1}{1+\frac{\Pi
(E)}{E-M_{0}+i\varepsilon}}\nonumber\\
&  =\frac{1}{E-M_{0}+\Pi(E)+i\varepsilon}\text{ .}\label{gssum}%
\end{align}
%

\begin{figure}
[ptb]
\begin{center}
\includegraphics[
height=1.8464in,
width=6.2119in
]%
{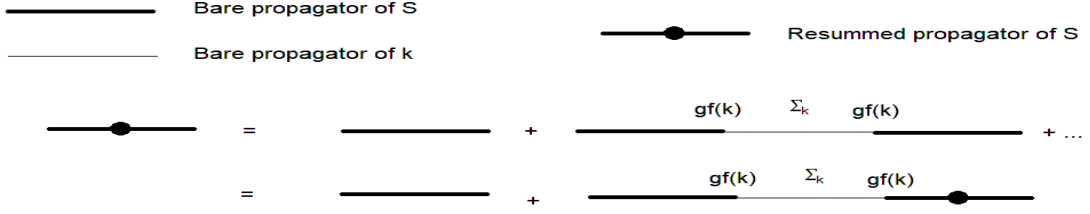}%
\caption{Schematic presentation of the sum leading to the dressed propagator.
In the last part the Bethe-Salpeter resummation is depicted. }%
\end{center}
\end{figure}
The sum in Eq. (\ref{gssum}) is shown in Fig. 1. It is interesting to notice
that the very same result can be obtained in an elegant way by using the
Bethe-Salpeter formalism \cite{bs} (see also Fig. 1):
\begin{equation}
G_{S}(E)=\frac{1}{E-M_{0}+i\varepsilon}-\frac{1}{E-M_{0}+i\varepsilon}%
\Pi(E)G_{S}(E)
\end{equation}
hence
\begin{equation}
G_{S}(E)\left(  1+\frac{\Pi(E)}{E-M_{0}+i\varepsilon}\right)  =\frac
{1}{E-M_{0}+i\varepsilon}%
\end{equation}
then%
\begin{equation}
G_{S}(E)=\frac{1}{E-M_{0}+\Pi(E)+i\varepsilon}\text{ .}%
\end{equation}

At this point we can identify Feynman rules for the Lee model:%
\begin{align}
\text{bare }S\text{ propagator } &  \rightarrow\frac{1}{E-M_{0}+i\varepsilon
}\text{ ;}\\
\text{bare }k\text{ propagator (}k\text{ fixed)} &  \rightarrow\frac
{1}{E-\omega(k)+i\varepsilon}\text{ ;}\\
kS\text{ vertex} &  \rightarrow gf(k)\text{ ;}\\
\text{internal }k\text{ line (}k\text{ not fixed)} &  \rightarrow-\Pi
(E)=\int_{-\infty}^{+\infty}\frac{dk}{2\pi}\frac{g^{2}f(k)^{2}}{E-\omega
(k)+i\varepsilon}\text{ .}%
\end{align}
Note, the latter expression can be understood as resulting from $gf(k)$ at
each vertex, by the $k$-propagator in the middle, and by an overall
integration $\int_{-\infty}^{+\infty}\frac{dk}{2\pi}$ due to the fact that $k$
is not fixed (just as a loop in QFT).

Finally, the survival amplitude can be expressed as%
\begin{equation}
a_{S}(t)=\frac{i}{2\pi}\int_{-\infty}^{+\infty}dEG_{S}(E)e^{-iEt}=\frac
{i}{2\pi}\int_{-\infty}^{+\infty}dE\frac{1}{E-M_{0}+\Pi(E)+i\varepsilon
}e^{-iEt}\text{ .}%
\end{equation}
This expression is an important intermediate result for the study of time
evolution of the unstable state $S,$ but it is not yet in the desired form of
Eq. (\ref{at}). In order to achieve that, additional steps are required.

\subsection{Definition of the spectral function}

Let us denote the basis of eigenstates of the Hamiltonian $H$ as $\left\vert
m\right\rangle $ with%
\begin{equation}
H\left\vert m\right\rangle =m\left\vert m\right\rangle \text{ for }m\geq
m_{th}\text{ (}m_{th}\text{ is the low-energy threshold) .}%
\end{equation}
The existence of a minimal energy $m_{th}$ is a general physical requirement.
The states $\left\vert m\right\rangle $ form an orthonormal basis of the
Hilbert space%
\begin{equation}
\mathcal{H=}\{\left\vert m\right\rangle \text{ with }m\geq m_{th}\}\text{ .}%
\end{equation}
The standard relations hold:
\begin{equation}
1_{\mathcal{H}}=\int_{m_{th}}^{+\infty}dm\left\vert m\right\rangle
\left\langle m\right\vert \text{ ,}%
\end{equation}%
\begin{equation}
\left\langle m_{1}|m_{2}\right\rangle =\delta(m_{1}-m_{2})\text{ .}%
\end{equation}

The link between the old basis $\{\left\vert S\right\rangle ,\left\vert
k\right\rangle \}$ (the eigenstates of $H_{0}$) and the new basis
$\{\left\vert m\right\rangle \}$ (he eigenstates of $H$) is not trivial. Yet,
for our purposes, the only requirement is that this basis of eigenstates of
$H$ exists. Indeed, this property follows from the fact that the Hamilton
operator is Hermitian.

Now, the state $\left\vert S\right\rangle $ can be expressed in terms of the
basis $\{\left\vert m\right\rangle \}$ as
\begin{equation}
\left\vert S\right\rangle =\int_{m_{th}}^{\infty}\alpha_{S}(m)\left\vert
m\right\rangle \text{ with }\alpha_{S}(m)=\left\langle S|m\right\rangle \text{
.}%
\end{equation}
The quantity
\begin{equation}
d_{S}(m)=\left\vert \alpha_{S}(m)\right\vert ^{2}=\left\vert \left\langle
S|m\right\rangle \right\vert ^{2}%
\end{equation}
is called the \textbf{spectral function (or energy/mass distribution) }of the
state $S$ (in agreement with the heuristic discussion of Sec. 2).

The normalization of the state $\left\vert S\right\rangle $ implies the
normalization of $d_{S}(m)$:
\begin{equation}
1=\left\langle S|S\right\rangle =\int_{m_{th}}^{\infty}d_{S}(m)dm\text{ .}%
\end{equation}
The simple intuitive interpretation is that $d_{S}(m)dm$ represents the
probability that the state $S$ has an energy (or mass) between $m$ and $m+dm.
$

As a consequence, the time-evolution can be easily evaluated by inserting
$1_{\mathcal{H}}=\int_{-\infty}^{+\infty}dm\left\vert m\right\rangle
\left\langle m\right\vert $ two times:%
\begin{equation}
a_{S}(t)=\left\langle S\right\vert U(t)\left\vert S\right\rangle =\left\langle
S\right\vert e^{-iHt}\left\vert S\right\rangle =\int_{m_{th}}^{\infty}%
d_{S}(m)e^{-imt}\text{ .}%
\end{equation}
This is all general, nice, and beautiful, but does not help us further as long
as we do not know how to calculate $d_{S}(m)$ in the framework of the Lee
model (or of any other model that we might use). This is fortunately possible
by using the propagator%

\begin{equation}
G_{S}(E)=\frac{1}{E-M_{0}+\Pi(E)+i\varepsilon}%
\end{equation}
which can be re-expressed as (inserting $1_{\mathcal{H}}=\int_{m_{th}%
}^{+\infty}dm\left\vert m\right\rangle \left\langle m\right\vert $ two
times):
\begin{align}
G_{S}(E) &  =\left\langle S\right\vert \frac{1}{E-H+i\varepsilon}\left\vert
S\right\rangle =\left\langle S\right\vert 1_{\mathcal{H}}\frac{1}%
{E-H+i\varepsilon}1_{\mathcal{H}}\left\vert S\right\rangle \nonumber\\
&  =\int_{m_{th}}^{\infty}dm_{1}\int_{m_{th}}^{\infty}dm_{2}\left\langle
m_{1}\right\vert \frac{1}{E-H+i\varepsilon}\left\vert m_{2}\right\rangle
\alpha_{S}^{\ast}(m_{1})\alpha_{S}(m_{2})\nonumber\\
&  =\int_{m_{th}}^{+\infty}dm\frac{d_{S}(m)}{E-m+i\varepsilon}\text{ .}%
\end{align}
Then, we obtain
\begin{equation}
G_{S}(E)=\frac{1}{E-M_{0}+\Pi(E)+i\varepsilon}=\int_{m_{th}}^{+\infty}%
dm\frac{d_{S}(m)}{E-m+i\varepsilon}\text{ .}\label{defdse}%
\end{equation}

Eq. (\ref{defdse}) can be considered as the definition of the spectral
function $d_{S}(m)$. Its physical meaning can be understood by noticing that
the dressed propagator $G_{S}(E)$ has been rewritten as the `sum' of free
propagators, whose weight function is $d_{S}(m).$ (In fact, $\left\vert
S\right\rangle $ is not an eigenstate of $H$ as soon as $H_{1}\neq0$).
Moreover, we expect that $d_{S}(m)\geq0$ and that the normalization
\begin{equation}
\int_{-\infty}^{+\infty}dmd_{S}(m)=\int_{m_{th}}^{+\infty}dmd_{S}(m)=1
\end{equation}
holds. The proof of the latter is indeed not trivial when starting from Eq.
(\ref{defdse}) (see the next subsection), but its physical and intuitive
justifications should be evident.

Let us turn to the evaluation of $d_{S}(E).$ First, let us consider the case
$g=0.$ In this limit, it is clear that:
\begin{equation}
d_{S}(E)=\delta(E-M_{0})\text{ .}%
\end{equation}
Namely, if the state $\left\vert S\right\rangle $ is an eigenstate of the
Hamiltonian, the mass distribution is a delta-function peaked at $M_{0}.$

When the interaction is switched on, we evaluate the imaginary part of Eq.
(\ref{defdse}):
\begin{equation}
\operatorname{Im}G_{S}(E)=\int_{-\infty}^{+\infty}dm\frac{-\varepsilon
}{(E-m)^{2}+\varepsilon^{2}}d_{S}(m)=-\int_{-\infty}^{+\infty}dm\pi
\delta(E-m)d_{S}(m)=-\pi d_{S}(E).
\end{equation}
Hence:%
\begin{equation}
d_{S}(E)=-\frac{\operatorname{Im}G_{S}(E)}{\pi}=\frac{1}{\pi}\frac
{\operatorname{Im}\Pi(E)}{(E-M_{0}+\operatorname{Re}\Pi(E))^{2}+\left(
\operatorname{Im}\Pi(E)\right)  ^{2}}\text{ .}\label{dse}%
\end{equation}

Once the spectral function $d_{S}(E)$ is known, the survival amplitude can be
expressed as its Fourier transform:%
\begin{align}
a_{S}(t)  &  =\frac{i}{2\pi}\int_{-\infty}^{+\infty}dEG_{S}(E)e^{-iEt}%
=\frac{i}{2\pi}\int_{-\infty}^{+\infty}dE\int_{-\infty}^{+\infty}dm\frac
{d_{S}(m)}{E-m+i\varepsilon}e^{-iEt}\nonumber\\
&  =\int_{-\infty}^{+\infty}d_{S}(m)dm\frac{i}{2\pi}\int_{-\infty}^{+\infty
}dE\frac{1}{E-m+i\varepsilon}e^{-iEt}=\int_{-\infty}^{+\infty}dmd_{S}%
(m)e^{-imt}\nonumber\\
&  =\int_{-\infty}^{+\infty}dEd_{S}(E)e^{-iEt}.
\end{align}

This is Eq. (\ref{at}) what we wanted to show: q.e.d.

\subsection{Proof of the normalization of the spectral function}

We now prove that the spectral function $d_{S}(E)$ calculated through Eq.
(\ref{dse}) is correctly normalized to $1$. Of course, this is compelling
since $a_{S}(0)=1$ is the starting point of our analysis. Yet, the
mathematical proof presented below (and based on Ref. \cite{lupoprd}) requires
some care.

First, we note that a low-energy threshold $m_{th}$ (hence a minimal energy)
is present in all physical system%
\begin{equation}
\operatorname{Im}\Pi(E)=0\text{ for }E<m_{th}\text{ }%
\end{equation}
($th$ stays for threshold).

Then, we first show the normalization under the assumption of a `strong'
requirement:%
\begin{equation}
\operatorname{Im}\Pi(E)=0\text{ for }E>\Lambda\text{ .}\label{strong}%
\end{equation}
This is not valid in general, but it allows for a simpler proof of the
normalization of $d_{S}(E).$ The real part of the loop $\operatorname{Re}%
\Pi(E)$ can be calculated from the dispersion relation%
\begin{equation}
\operatorname{Re}\Pi(E)=\frac{1}{\pi}PP\int_{m_{th}}^{\infty}\frac
{\operatorname{Im}\Pi(m)}{E-m}dm=\frac{1}{\pi}PP\int_{m_{th}}^{\Lambda}%
\frac{\operatorname{Im}\Pi(m)}{E-m}dm\text{ ,}%
\end{equation}
where $PP$ stays for principal part, out of which one can see that
$\operatorname{Re}\Pi(E)$ goes to zero as $1/E$ for $E\gg\Lambda$. Hence,
taking the limit $E\rightarrow\infty,$ one gets%

\begin{align}
\lim_{E\rightarrow\infty}\frac{1}{E-M_{0}+\Pi(E)+i\varepsilon}  &  =\frac
{1}{E}\\
&  =\lim_{E\rightarrow\infty}\int_{m_{th}}^{\Lambda}dm\frac{d_{S}%
(m)}{E-m+i\varepsilon}=\frac{1}{E}\int_{m_{th}}^{\Lambda}dmd_{S}(m)
\end{align}
Hence,
\begin{equation}
\int_{E_{th}}^{\Lambda}dmd_{S}(m)=1
\end{equation}
follows.

Now, we release the `strong' assumption (\ref{strong}), but we assume that
$\operatorname{Im}\Pi(E)$ goes to zero sufficiently fast as function of $E$
for $E\rightarrow\infty$. Then, we rewrite:
\begin{equation}
\int_{m_{th}}^{\infty}dm\frac{d_{S}(m)}{E-m+i\varepsilon}=\int_{m_{th}}%
^{\sqrt{M_{0}E}}dm\frac{d_{S}(m)}{E-m+i\varepsilon}+\int_{\sqrt{M_{0}E}%
}^{\infty}dm\frac{d_{S}(m)}{E-E^{^{\prime}}+i\varepsilon}=I_{1}+I_{2}\text{ .}%
\end{equation}
We have divided the integral into two pieces by setting the division at
$\sqrt{M_{0}E}.$ The result, of course, does not depend on this choice (if on
takes, for instance, $2\sqrt{M_{0}E},$ nothing changes). This separation is
useful. Namely, the large-$E$ limit of the first integral is easily taken,
because no pole is present in the integration (inf fact, $E$ is surely larger
than $\sqrt{M_{0}E}$ in the large-$E$ limit):
\begin{equation}
\lim_{E\rightarrow\infty}I_{1}(E)=\lim_{E\rightarrow\infty}\int_{m_{th}%
}^{\sqrt{M_{0}E}}dm\frac{d_{S}(m)}{E-m+i\varepsilon}=\frac{1}{E}\int_{m_{th}%
}^{\infty}dmd_{S}(m)
\end{equation}
Then, the second integral takes the form:
\begin{equation}
I_{2}(E)=\int_{\sqrt{M_{0}E}}^{\infty}dm\frac{d_{S}(m)}{E-m+i\varepsilon}%
=\int_{\sqrt{M_{0}E}}^{\infty}dm\frac{1}{E-m+i\varepsilon}\frac{1}{\pi}%
\frac{\operatorname{Im}\Pi(m)}{(m-M_{0}+\operatorname{Re}\Pi(m))^{2}+\left(
\operatorname{Im}\Pi(m)\right)  ^{2}}\text{ .}%
\end{equation}
It is then clear that $\operatorname{Im}I_{2}=d_{S}(E)$, which is very small
for large $E.$ Next, the real part of $I_{2}$ reads%
\begin{equation}
\operatorname{Re}I_{2}(E)=PP\int_{\sqrt{M_{0}E}}^{\infty}dm\frac{1}{\pi}%
\frac{1}{E-m+i\varepsilon}\frac{\operatorname{Im}\Pi(m)}{(m-M_{0}%
+\operatorname{Re}\Pi(E^{\prime}))^{2}+\left(  \operatorname{Im}\Pi(E^{\prime
})\right)  ^{2}}\text{ }.
\end{equation}
We assume that $\operatorname{Im}\Pi(m)$ goes to zero sufficiently fast for
$m\rightarrow\infty$ in such a way that $\operatorname{Re}I_{2}$ vanishes.
Then, one has:
\begin{equation}
\lim_{E\rightarrow\infty}I_{2}(E)=0.
\end{equation}
Hence:
\begin{equation}
\int_{m_{th}}^{\infty}dmd_{S}(m)=\int_{m_{th}}^{\infty}dEd_{S}(E)=\int
_{-\infty}^{+\infty}dEd_{S}(E)=1
\end{equation}
is proven, which corresponds to our second goal mentioned in the introduction,
the verification of Eq. (\ref{norm}): q.e.d.

\subsection{The Breit-Wigner limit}

As a last point we discuss the Breit-Wigner limit \cite{scully,ww}. To this
end, we use the LLM discussed in\ Sec. 3, $\omega(k)=k,$ together with the
modulation function%

\begin{equation}
f(k)=\theta(M_{0}+\Lambda-k)\theta(k-(M_{0}-\Lambda)).
\end{equation}
In this way, the unstable state $\left\vert S\right\rangle $ couples in a
limited $\emph{window}$ of energy to the final states of the type $\left\vert
k\right\rangle $ (see also \cite{pra1} for details).

The self-energy $\Sigma(E)$ reads%
\begin{equation}
\Sigma(E)=\frac{g^{2}}{2\pi}\ln\left(  \frac{E-M_{0}+\Lambda}{E-M_{0}-\Lambda
}\right)  \text{ ,}%
\end{equation}
whose real and imaginary parts are
\begin{align}
\operatorname{Re}\Sigma(E)  &  =\frac{g^{2}}{2\pi}\ln\left\vert \frac
{E-M_{0}+\Lambda}{E-M_{0}-\Lambda}\right\vert \text{ , }\\
\operatorname{Im}\Sigma(E)  &  =\left\{
\begin{array}
[c]{c}%
\frac{g^{2}}{2}\text{ for }M_{0}-\Lambda<E<M_{0}+\Lambda\\
0\text{ otherwise}%
\end{array}
\right.  .
\end{align}

When $\Lambda$ is not infinite, deviations both at short and long times occur.
Yet, in the limit $\Lambda\rightarrow\infty$ one recovers the pure exponential
decay. Namely:
\begin{equation}
\operatorname{Re}\Sigma(E)=0\text{ and }\operatorname{Im}\Sigma(E)=\frac
{g^{2}}{2}\text{for each }E\text{ .}%
\end{equation}
The propagator reduces exactly to the BW form
\begin{equation}
G_{S}(E)=\frac{1}{E-M_{\text{BW}}+i\Gamma_{\text{BW}}/2}%
\end{equation}
with
\begin{equation}
M_{\text{BW}}=M_{0}\text{ and }\Gamma_{\text{BW}}=g^{2}.
\end{equation}

The survival probability amplitude of the state $\left\vert S\right\rangle $
is also in this case the usual exponential form%
\begin{align}
a_{S}(t)  &  =\left\langle S\right\vert e^{-iHt}\left\vert S\right\rangle
=\frac{i}{2\pi}\int_{-\infty}^{+\infty}dEG_{S}(E)e^{-iEt}=\nonumber\\
&  =\frac{i}{2\pi}\int_{-\infty}^{+\infty}dE\frac{1}{E-M_{\text{BW}}%
+i\Gamma_{\text{BW}}/2}e^{-iEt}=e^{-i(M_{\text{BW}}-i\Gamma_{\text{BW}}/2)t}%
\end{align}
where the pole $E=M_{\text{BW}}-i\Gamma_{\text{BW}}/2$ is picked up when
performing the integration.

The spectral function takes the form%
\begin{equation}
d_{S}(E)=-\frac{\operatorname{Im}G_{S}(E)}{\pi}=\frac{\Gamma_{\text{BW}}}%
{2\pi}\frac{1}{(E-M_{\text{BW}})^{2}+\Gamma_{\text{BW}}^{2}/4}=d_{S}%
^{\text{BW}}(E)\text{ ,}%
\end{equation}
i.e. the usual Breit-Wigner form introduced already in the introduction. The
survival probability amplitude can be also calculated by using Eq. (\ref{at})
obtaining
\begin{equation}
a_{s}(t)=\left\langle S\right\vert e^{-iHt}\left\vert S\right\rangle
=\int_{-\infty}^{+\infty}dEd_{S}^{\text{BW}}(E)e^{-iEt}=e^{-i(M_{\text{BW}%
}-i\Gamma_{\text{BW}}/2)t}\text{ ,}%
\end{equation}
out of which
\begin{equation}
p_{S}(t)=e^{-\Gamma_{\text{BW}}t}\text{ }.
\end{equation}

Quite interestingly, in the BW case it is also possible to evaluate the
time-evolution operator applied to the unstable state $\left\vert
S\right\rangle $ \cite{pra1}:%
\begin{equation}
e^{-iHt}\left\vert S\right\rangle =e^{-i(M_{\text{BW}}-i\Gamma_{\text{BW}%
}/2)t}\left\vert S\right\rangle +\int_{-\infty}^{+\infty}dkb(k,t)\left\vert
k\right\rangle
\end{equation}
with%
\begin{equation}
b(k,t)=\frac{g}{\sqrt{2\pi}}\frac{e^{-ikt}-e^{-i(M_{\text{BW}}-i\Gamma
_{\text{BW}}/2)t}}{k-M_{\text{BW}}+i\Gamma_{\text{BW}}/2}\text{ .}%
\end{equation}
Obviously, the probability that the decay has occurred is
\begin{equation}
w(t)=\int_{-\infty}^{+\infty}dk\left\vert b(k,t)\right\vert ^{2}%
=1-p(t)=1-e^{-\Gamma_{\text{BW}}t}\text{ .}%
\end{equation}

In the end, note that in the BW limit we could describe the evolution of the
state $\left\vert S\right\rangle $ by an \textquotedblleft
effective\textquotedblright\ non-Hermitian Hamiltonian%
\begin{equation}
H_{eff,S}=\left(  M_{0}-i\frac{\Gamma}{2}\right)  \left\vert S\right\rangle
\left\langle S\right\vert \text{ .}%
\end{equation}
Yet, such as an expression -although useful in some cases- should be regarded
with due care.

Finally, we could show that under certain restrictive conditions the BW limit
is recovered. As a next step, we should discuss that the BW expressions
represent indeed a good approximations in most physical cases.

\subsection{Breit-Wigner approximation: mass and width}

Let us now consider the case in which we do not have exactly a BW spectral
function, but it is still possible to show how the latter emerges as an approximation.

Let us consider the propagator
\begin{equation}
G_{S}(E)=\frac{1}{E-M_{0}+\Pi(E)+i\varepsilon}=\frac{1}{E-M_{0}%
+\operatorname{Re}\Pi(E)+i\operatorname{Im}\Pi(E)+i\varepsilon}\text{ .}%
\end{equation}

The (renormalized) nominal BW\ mass of the state $\left\vert S\right\rangle $
is defined as the solution of the
\begin{equation}
M_{\text{BW}}-M_{0}+g^{2}\operatorname{Re}\Sigma(M_{\text{BW}})=0\text{ .}%
\end{equation}
By expanding the real part of $G_{S}^{-1}(E)$ around $M_{\text{BW}},$ we
obtain
\begin{align}
G_{S}(E)  &  =\frac{1}{E-M_{\text{BW}}+\Pi(E)+i\varepsilon}\nonumber\\
&  =\frac{1}{(E-M_{\text{BW}})\left(  1+g^{2}\left(  \frac{\partial
\operatorname{Re}\Sigma(E)}{\partial E}\right)  _{E=M}+...\right)
+i\operatorname{Im}\Pi(E)+i\varepsilon}\nonumber\\
&  \simeq\frac{1}{\left(  1+g^{2}\left(  \frac{\partial\operatorname{Re}%
\Sigma(E)}{\partial E}\right)  _{E=M_{\text{BW}}}\right)  }\frac
{1}{E-M_{\text{BW}}+i\Gamma_{\text{BW}}}%
\end{align}

Hence, the Breit-Wigner approximation of the propagator emerges as
\begin{equation}
G_{S}^{\text{BW}}(E)=Z_{\text{BW}}\frac{1}{E-M+i\Gamma_{\text{BW}}%
/2}\label{gsbw}%
\end{equation}
where
\begin{equation}
Z_{\text{BW}}=\left(  1+g^{2}\left(  \frac{\partial\operatorname{Re}\Sigma
(E)}{\partial E}\right)  _{E=M_{\text{BW}}}\right)  ^{-1}%
\end{equation}
is the normalization constant. The decay width $\Gamma_{\text{BW}}$ is given
by (an extension of) the Fermi golden rule
\begin{equation}
\Gamma_{\text{BW}}=\frac{g^{2}}{1+g^{2}\left(  \frac{\partial\operatorname{Re}%
\Sigma(E)}{\partial E}\right)  _{E=M_{\text{BW}}}}\operatorname{Im}%
\Sigma(M_{\text{BW}})=\frac{g^{2}}{1+g^{2}\left(  \frac{\partial
\operatorname{Re}\Sigma(E)}{\partial E}\right)  _{E=M}}\frac{f^{2}(k_{M}%
)}{\left(  \frac{d\omega}{dk}\right)  _{k=k_{M}}}\text{ }%
\end{equation}
where $k_{M}$ given by
\begin{equation}
\omega(k_{M})=M_{\text{BW}}\text{ .}%
\end{equation}
Then, using the approximation in Eq. (\ref{gsbw}), the survival probability
amplitude is given by
\begin{equation}
a_{S}(t)=\left\langle S\right\vert e^{-iHt}\left\vert S\right\rangle
=e^{-i(M_{\text{BW}}-i\Gamma_{\text{BW}}/2)t}%
\end{equation}
and the survival probability takes the usual form $p_{S}(t)=\left\vert
a_{S}(t)\right\vert ^{2}\simeq\left\vert Z_{\text{BW}}\right\vert
^{2}e^{-\Gamma_{\text{BW}}t}$. One can then also see that the exponential
limit is recovered, but there is a constant $\left\vert Z_{\text{BW}%
}\right\vert ^{2}$ which differs from $1$ in front of it (this fact also
implies that for short times deviations from the exponential decay are
present; for a detailed discussion of this point see Ref.
\cite{fptopicalreview}).

Very often (see e.g. Ref. \cite{pelaezrev} and refs. therein) one extends the
propagator to the complex plane upon considering $E\rightarrow z\in%
\mathbb{C}
$:
\begin{equation}
G_{S}(z)=\frac{1}{z-M_{0}+\Pi(z)}%
\end{equation}
where the loop function on the complex plane reads
\begin{equation}
\Pi(z)=\frac{1}{\pi}\int_{m_{th}}^{\infty}\frac{\operatorname{Im}\Pi(m)}%
{m-z}dm\text{ .}%
\end{equation}

Next, one searches for the pole(s) of $G_{S}(z)$ in the complex plane in the
II-Riemann sheet
\begin{equation}
z_{\text{pole}}-M_{0}+\Pi_{II}(z_{\text{pole}})=0\text{ }%
\end{equation}
where the loop on the second Riemann sheet is given by:%
\begin{equation}
\Sigma_{II}(z)=\Sigma(z)+2i\operatorname{Im}\Sigma(z)\text{ }%
\end{equation}
with $\operatorname{Im}\Sigma(z)$ being the imaginary part of the loop
analytically continued to the whole complex plane.

Typically, there is one dominating pole close to the real axis, for which the
mass $M_{\text{pole}}$ and the width $\Gamma_{\text{pole}}$ of the unstable
state are defined as%
\begin{equation}
z_{\text{pole}}=M_{\text{pole}}-i\Gamma_{\text{pole}}/2\text{ .}%
\end{equation}
When considering $z$ close to the pole one can write
\begin{equation}
G_{S}(z)\simeq\frac{Z_{\text{pole}}}{z-z_{\text{pole}}}%
\end{equation}
where $Z$ is the residue of the pole. The evaluation of the survival
probability amplitude under the assumption that a single pole dominates leads
to%
\begin{align}
a_{S}(t) &  =\left\langle S\right\vert e^{-iHt}\left\vert S\right\rangle
=\frac{i}{2\pi}\int_{-\infty}^{+\infty}dEG_{S}(E)e^{-iEt}=\nonumber\\
&  \simeq\frac{i}{2\pi}\int_{-\infty}^{+\infty}dE\frac{Z_{\text{pole}}%
}{E-z_{\text{pole}}}e^{-iEt}=Z_{\text{pole}}e^{-i(M_{\text{pole}}%
-i\Gamma_{\text{pole}}/2)t}%
\end{align}
hence $p_{S}(t)\simeq\left\vert Z_{\text{pole}}\right\vert ^{2}e^{-\Gamma
_{\text{pole}}t}.$ The form is identical to the BW one, but the numerical
results for masses and decays are not exactly equal (they coincide only in the
small-width limit).

The pole mass and width are typically preferable from a theoretical point of
view than the BW mass and width since the position of the pole is process
independent \cite{pelaezrev}. Yet, both of them are commonly used in practice
\cite{pdg}.

\section{Applications of the Lee model}

The Lee model has been commonly employed to describe various problems in
different area of physics. In connection to decays of quantum states and their
connection to the QZE and IZE effects, it has been used in e.g. Refs.
\cite{gadella,facchiprl,shimizu,kk1,kk2,fptopicalreview,fp} and references therein.

In Table 1 we present some additional recent studies that made use of the Lee
model in order to study related but somewhat different topics.

\begin{center}
Table 1: some selected recent studies using the Lee model%

\begin{tabular}
[c]{||l||l||}\hline\hline
\textbf{Topic} & \textbf{Ref}\\\hline\hline
Two-channel decay & \cite{duecan}\\\hline\hline
Moving unstable state, time dilation & \cite{acta,adv}\\\hline\hline
Delta resonance & \cite{delta}\\\hline\hline
$X(3872)$ & \cite{x3872}\\\hline\hline
Finite Temperature & \cite{pok}\\\hline\hline
Broadening of the spectrum & \cite{pra1}\\\hline\hline
QZE and IZE (and fundamental issues) & \cite{pra2}\\\hline\hline
\end{tabular}

\end{center}

Some comments are in order:

(i) In the first entry of Table 1 the extension to two decay channel is
mentioned. This is quite important since the majority of unstable states has
more than a single decay channel. The extension of the Lee model in this case
is simple: we couple the state $\left\vert S\right\rangle $ to two sets of
final states $\left\vert k,1\right\rangle $ and $\left\vert k,2\right\rangle
.$ The Hamiltonian reads%

\begin{equation}
H_{0}=M\left\vert S\right\rangle \left\langle S\right\vert +\sum_{i=1,2}%
\int_{-\infty}^{+\infty}dk\omega_{i}(k)\left\vert k,i\right\rangle
\left\langle k,i\right\vert \text{ , }H_{1}=\sum_{i=1,2}\int_{-\infty
}^{+\infty}dk\frac{g_{i}f_{i}(k)}{\sqrt{2\pi}}\left(  \left\vert
S\right\rangle \left\langle k,i\right\vert +\text{h.c.}\right)  \text{
.}\label{duecan}%
\end{equation}
In particular, the BW limit in the LLM (by repeating the steps of Sec. 4.5)
implies that the partial decay widths are $\Gamma_{1}=g_{1}^{2}$ and
$\Gamma_{2}=g_{2}^{2}$ and the survival probability reads $p_{S}%
(t)=e^{-\Gamma_{\text{BW}}t}$ with $\Gamma_{\text{BW}}=\Gamma_{1}+\Gamma_{2}$.

In the presence of two decay channels, it is useful to introduce the quantity
$h_{i}(t)$: $h_{i}(t)dt$ is the probability that the state $\left\vert
S\right\rangle $ decays in the $i$-th channel between $t$ and $t+dt.$ In the
BW limit $h_{i}(t)=\Gamma_{i}e^{-\Gamma t}$ and the ratio $h_{1}(t)/h_{2}(t)$
is a constant equal to $\Gamma_{1}/\Gamma_{2}$. However, when deviations from
the exponential decay are considered, the ratio $h_{1}(t)/h_{2}(t)$ is in
general not a constant, but shows sizable departures from $\Gamma_{1}%
/\Gamma_{2}$ \cite{duecan}. In fact, this ratio presents large and irregular
oscillations which persist for a long time, even in the regime in which the
decay law $p_{S}(t)$ is very well approximated by an exponential function. In
this sense, this ratio represents a novel tool to detect deviations from the
exponential decay. For a very recent discussion of this problem by using a QM
model, see Ref. \cite{koscik}.

(ii) The decay of the unstable $S$ is calculated -as usual- in the rest frame
of the particle $S.$ A very interesting question is the evaluation of the
survival probability in the case in which the particle is moving. By denoting
$p$ as the modulus of the three-momentum of $S$, one obtains%
\begin{equation}
a_{S}^{p}(t)=\int_{-\infty}^{+\infty}d_{S}(E)e^{-i\sqrt{E^{2}+p^{2}}t}dE\text{
}.
\end{equation}
For the derivation of this result by using the Lee model see Refs.
\cite{acta,adv}. (For the discussion of this topic, see also Refs.
\cite{khalfin2,shirokov,stefanovich,urbanowski}.) In particular, one finds
that the usual dilation formula is not reobtained. in the BW limit one finds
that
\begin{equation}
\left\vert a_{S}^{p}(t)\right\vert ^{2}\neq e^{-\frac{\Gamma}{\gamma}t}%
\end{equation}
where $\gamma=\sqrt{p^{2}+M_{\text{BW}}^{2}}/M_{\text{BW}}$ is the Lorentz
factor (for the explicit expression of $p_{S}(t)$ in this case, see Ref.
\cite{acta}). Obviously, $e^{-\frac{\Gamma}{\gamma}t}$ is normally used in
practice. Indeed, very small deviations from it are present. It should be also
stressed that there is no violation of special relativity but that care is
needed when an unstable state with nonzero momentum is defined, see details
in\ Ref. \cite{acta}.

(iii-iv) In the third and fourth entries of Table 1, two resonances are
mentioned: the baryon $\Delta$ \cite{delta} and the enigmatic $X(3872)$ state
\cite{x3872} (for the role of loops in the latter see also Ref. \cite{coitox}%
). In general, one can use similar techniques for any resonance.

(v) Extension to finite temperature. The Lee model can be also used at finite
temperature in order to study how to threat unstable resonances in a thermal
gas. This has been recently accomplished in\ Ref. \cite{pok} where the
so-called `phase-shift' formula for the proper description of resonances at a
given temperature could be proven to be exact within the Lee model.

(vi-vii) The Lee model has been utilized in Ref. \cite{pra1} to study the
broadening of the energy spectrum of an unstable state if the measurement is
performed early enough. In a further extension, a discussion of fundamental
properties -such as the Zeno effect induced by imperfect measurements and the
possible connection to the Many World Interpretation of QM has been discussed
in\ Ref. \cite{pra2}.

Finally,a consideration concerning the connection of the lee model to QFT is
necessary. In Ref. \cite{duecan} the comparison of the Lee model with QFT
approaches is presented. In particular, it is shown that the QFT counterpart
is given by the interaction Lagrangian
\begin{equation}
\mathcal{L}=gS\varphi^{2}\text{ ,}%
\end{equation}
which describes the two-decay process $S\rightarrow\varphi\varphi$ (see
\cite{lupo} for technical details of the QFT treatment). Hence, the field $S$
corresponds the ket $\left\vert S\right\rangle $ described by the Lee
Hamiltonian and the two-state $\varphi\varphi$ corresponds to $\left\vert
k\right\rangle .$ It is then also possible to verify that deviations form the
exponential decay are realized in QFT as well \cite{zenoqft}. Yet, even if the
Lee model presents many features of QFT, it is not QFT. The fact is that in a
genuine QFT approach also transitions of the type $S\varphi^{2}\equiv
\left\vert Sk\right\rangle \rightarrow\left\vert 0\right\rangle $ (the
perturbative vacuum) and vice-versa are possible, which are however not
included in the Lee model. Moreover, QFT allows for an arbitrary number of $S
$ and $\varphi$ fields (and not necessarily $1$ and $2,$ as in the Lee
approach). An additional subtle but important problem concerns the
identification of the real vacuum of the theory (which is not the perturbative
vacuum) \cite{peskin,blasone}, which is necessary for a proper introduction of
an unstable state in QFT. This is indeed an interesting topic for future developments.

\section{Conclusions}

In this work we have described the Lee model by paying attention to many
technical details. To this end, we have introduced it as a limiting process of
a discrete Lee model. We have shown how the survival decay amplitude can be
properly derived within this framework as the Fourier transform of the
spectral function. The latter emerges as the imaginary part of the propagator
of the unstable state under study and turns out to be normalized to unity, as
we have proven by a detailed analysis.

Moreover, we have also shown how the BW limit emerges as a particular
approximation of the spectral function and how the BW mass and widths are
correctly defined.\ In addition, the pole and mass and width have been also be introduced

The Lee model is a very versatile approach that can be used to test and
discuss many different physical situations which ranges from QM systems to
purely QFT ones, as we have illustrated in\ Sec. 6. Hopefully, the detailed
presentation of this work may help to initiate new studies that make use of
this useful and beautiful model in various areas of physics.

\bigskip

\textbf{Acknowledgments: } the author thanks G. Pagliara and S.
Mr\'{o}wczy\'{n}sky for useful discussions.

\end{document}